\documentclass[a4paper,reqno,12pt,final]{amsart}
\usepackage{amsmath,amsfonts,amssymb,amsthm,amstext,amscd,array,bbold}
\usepackage[latin1]{inputenc}
\DeclareMathOperator{\AdS}{AdS}
\DeclareMathOperator{\dS}{dS}
\DeclareMathOperator{\Aut}{Aut}
\DeclareMathOperator{\Out}{Out}
\DeclareMathOperator{\Inn}{Inn}
\DeclareMathOperator{\Tr}{Tr}
\DeclareMathOperator{\image}{im}
\DeclareMathOperator{\id}{id}
\DeclareMathOperator{\im}{Im}
\DeclareMathOperator{\re}{Re}
\DeclareMathOperator{\NW}{NW}
\DeclareMathOperator{\II}{II}
\newcommand{\CC}{\mathbb{C}}
\newcommand{\RR}{\mathbb{R}}
\newcommand{\ZZ}{\mathbb{Z}}
\newcommand{\EE}{\mathbb{E}}
\newcommand{\fsu}{\mathfrak{su}}
\newcommand{\fso}{\mathfrak{so}}
\newcommand{\fn}{\mathfrak{n}}
\newcommand{\half}{\tfrac{1}{2}}
\newcommand{\SU}{\mathrm{SU}}
\newcommand{\SO}{\mathrm{SO}}
\newcommand{\GL}{\mathrm{GL}}
\renewcommand{\O}{\mathrm{O}}
\newcommand{\U}{\mathrm{U}}
\newcommand{\sX}{\mathsf{X}}
\newcommand{\sU}{\mathsf{U}}
\newcommand{\sV}{\mathsf{V}}
\newcommand{\sP}{\mathsf{P}}
\newcommand{\sJ}{\mathsf{J}}
\newcommand{\sK}{\mathsf{K}}
\newcommand{\eN}{\mathcal{N}}
\newcommand{\eC}{\mathcal{C}}
\newcommand{\eZ}{\mathcal{Z}}
\newcommand{\bx}{\boldsymbol{x}}
\newcommand{\by}{\boldsymbol{y}}
\newcommand{\1}{\mathbb{1}}
\allowdisplaybreaks[1]
\begin{document}
\title[Penrose limits and a Nappi--Witten braneworld]{Penrose limits
  of Lie Branes and a Nappi--Witten braneworld}
\author[Stanciu]{Sonia Stanciu$^\maltese$}
\address{Theoretical Physics Group, Imperial College, London, UK}
\author[Figueroa-O'Farrill]{José Figueroa-O'Farrill}
\address{Department of Mathematics, University of Edinburgh, UK}
\email{j.m.figueroa@ed.ac.uk}
\thanks{EMPG-03-07}
\begin{abstract}
  Departing from the observation that the Penrose limit of $\AdS_3
  \times S^3$ is a group contraction in the sense of Inönü and Wigner,
  we explore the relation between the symmetric D-branes of $\AdS_3
  \times S^3$ and those of its Penrose limit, a six-dimensional
  symmetric plane wave analogous to the four-dimensional Nappi--Witten
  spacetime.  Both backgrounds are Lie groups admitting bi-invariant
  lorentzian metrics and symmetric D-branes wrap their (twisted)
  conjugacy classes.  We determine the (twisted and untwisted)
  symmetric D-branes in the plane wave background and we prove the
  existence of a space-filling D5-brane and, separately, of a
  foliation by D3-branes with the geometry of the Nappi--Witten
  spacetime which can be understood as the Penrose limit of the
  $\AdS_2 \times S^2$ D3-brane in $\AdS_3 \times S^3$.
  Parenthetically we also derive a simple criterion for a symmetric
  plane wave to be isometric to a lorentzian Lie group.  In particular
  we observe that the maximally supersymmetric plane wave in IIB
  string theory is isometric to a lorentzian Lie group, whereas the
  one in M-theory is not.
\end{abstract}
\maketitle
\tableofcontents

\section{Introduction}

The Penrose limit \cite{PenrosePlaneWave, GuevenPlaneWave}, originally
a curiosity of four-dimensional General Relativity without many
practical applications, has attracted much attention recently, largely
due to its role in the novel large $N$ limit of supersymmetric gauge
theories discovered in \cite{MaldaPL} and vigorously studied since.
The classical ``pointlike'' geometry of the Penrose limit is by now
well understood \cite{Limits}, but strings probe other aspects of the
spacetime geometry (e.g., D-submanifolds).  It is therefore a natural
question to ask how these behave under the Penrose limit.

This project thus originated as an attempt to understand what the
Penrose limit, in its more recent stringy avatar, does to branes.  In
other words, if two string backgrounds are related by a Penrose limit,
then how, if in any way, do the branes in one background relate to the
branes in the other background?  A priori one might suspect that there
is little or no relation, since a brane might be localised far away
from the null geodesic along which we are performing the Penrose
limit, and hence will be lost in the limit.  On the other hand, it is
always possible to consider lorentzian branes which intersect a null
geodesic and see if and how the limit of the background induces a
limit of the brane.

Lie groups admitting a bi-invariant metric comprise a class of exact
string backgrounds whose D-submanifolds (at least the symmetric ones)
are well understood geometrically.  It is therefore reasonable to
investigate these backgrounds first and study what happens to Lie
branes\footnote{Throughout this article we use the term ``Lie brane''
  as a shorthand for ``symmetric D-brane in a Lie group.''}  in the
limit.  In this paper we will collect our initial results in this
topic by studying two vacua of the minimal chiral six-dimensional
supergravity.  These backgrounds can be lifted to exact backgrounds of
ten-dimensional string theory by crossing them with some
four-dimensional exact background like flat space or a K3 surface, for
instance.  However we will ignore the four-dimensional factor in what
follows and concentrate on the six-dimensional geometry.

The backgrounds in question are $\AdS_3 \times S^3$ and the
six-dimensional version $\NW_6$ of the Nappi--Witten spacetime
\cite{NW} discovered in \cite{Meessen} in the supergravity context and
denoted there $\text{KG}_6$.  Both of these backgrounds are Lie groups
and the Penrose limit which relates them can also be interpreted as a
group contraction in the sense of Inönü--Wigner \cite{InonuWigner}.
This will be explained in Section~\ref{sec:PL=IWC}, which also
contains a subsection on the geometric characterisation of those
symmetric plane waves which are isometric to Lie groups admitting a
bi-invariant metric.  Strings propagating in such Lie groups are
described by a WZW model, whose symmetric D-submanifolds are known to
be given by (shifted, twisted) conjugacy classes
\cite{AS,FFFS,SDnotes,FSNW}.  The symmetric D-submanifolds for $\AdS_3
\times S^3$ were determined in \cite{Sads3,BPAdS2} and the results are
reviewed in Section~\ref{sec:LBAdSxS} to ease the comparison.  Those
in $\NW_6$ are worked out in Section~\ref{sec:LBNW6} for the first
time following the method of \cite{FSNW} for the four-dimensional
Nappi--Witten spacetime $\NW_4$.  In particular we demonstrate the
existence of D3-branes isometric to $\NW_4$ which foliate $\NW_6$ in a
variety of ways.  This is done in two ways: by a geometric argument
and by an explicit calculation.  We also show that $\NW_6$ admits a
space-filling Lie brane.  Both are examples of twisted conjugacy
classes by a nontrivial outer automorphism.

\section*{Note added}

\begin{flushright}
  \begin{scriptsize}
    \textsl{¿Y ha de morir contigo el mundo tuyo,\\
      la vieja vida en orden tuyo y nuevo?\\
      ¿Los yunques y crisoles de tu alma\\
      trabajan para el polvo y para el viento?}\\
  
    \hfill Antonio \textsc{Machado}
\end{scriptsize}
\end{flushright}

Most of the work contained in this paper was finished in March 2002,
weeks before the terrible illness which so tragically cut short the
life of the first named author began to take its toll.  The decision
to finally write up the paper at this late stage is due in part to the
fact that some of the results presented here are revisited in
forthcoming work involving the second named author, who would like to
take this opportunity to apologise for the delay in completing the
present paper.

In the intervening time a number of papers have appeared dealing with
D-branes in the Penrose limit of $\AdS_3 \times S^3$ in the
supergravity approximation.  Two papers \cite{BKPDpp,NayakDangles}
deal with intersecting branes (at angles).  As remarked at the end of
\cite[Section~2.3]{FSNW}, Lie branes always wrap \emph{submanifolds},
whence intersecting brane configurations are not described by twisted
conjugacy classes, at least in a straightforward manner.  Our results
therefore bear no comparison.  Three other papers can be compared and
agree with our results.  In \cite{KNSD5} the authors construct D5- and
NS5-brane solutions of type II supergravity whose worldvolume is the
space-filling brane in $\NW_6$ described in Section~\ref{sec:D5} of
the present paper.  These solutions are obtained as the Penrose limits
of fivebranes in $\AdS_3 \times S^3 \times \RR^4$ with worldvolumes
which are space-filling in $\AdS_3 \times S^3$.  These fivebranes are
not Lie branes, though.  In \cite{Singh}, the authors constructs an
M5-brane solution whose worldvolume geometry agrees with the
space-filling brane above by taking a Penrose limit of an M5-solution
with worldvolume geometry $\AdS_3 \times S^3$.  Finally, in
\cite{AKD3} the authors find, in the T-duality orbit of the D5-brane
solution in \cite{KNSD5}, a D3-brane solution whose worldvolume is
isometric to the Nappi--Witten spacetime $\NW_4$ and hence to the
symmetric D3-branes described in Section~\ref{sec:D3}.  The
Nappi--Witten spacetime also arises as a Penrose limit of the near
horizon geometry of a NS5-brane \cite{GONW4}.

\section{The Penrose limits of $\AdS_3 \times S^3$ as group
  contraction}
\label{sec:PL=IWC}

The Penrose limits of $\AdS_3 \times S^3$ were recently discussed in
\cite{ShortLimits,Limits}.  There are two possible limits: the generic
Penrose limit is a lorentzian symmetric space which, if the radii of
curvature of $\AdS_3$ and $S^3$ are the same, is conformally flat.
There is a also a special limit in which the resulting spacetime is
flat.  In the generic case with equal radii of curvature, we can also
understand this Penrose limit by isometrically embedding $\AdS_3
\times S^3$ in $\RR^8$ with a flat metric of signature $(6,2)$.  The
Penrose limit is then induced by a generalised Penrose limit along a
null plane through the origin, as discussed in \cite{Limits}.

\subsection{Penrose limit as a group contraction}

In this section we will discuss the Penrose limit in terms of group
contraction.  The manifold $\AdS_3 \times S^3$ is isometric to the Lie
group $\SU(1,1) \times \SU(2)$ with a bi-invariant metric.
One-parameter subgroups are geodesic relative to bi-invariant metrics,
hence any one-parameter subgroup which is null relative to the
bi-invariant metric gives rise to a null geodesic whose associated
Penrose limit can be understood as a group contraction.  Such special
cases of Penrose limits have been considered in \cite{ORS}.

We will identify the Lie groups $\SU(1,1)$  and $\SU(2)$ with their
defining representations.  That means that $\SU(1,1)$ is the group of
matrices of the form
\begin{equation}
  \label{eq:su11}
  \begin{pmatrix}
    a & b \\ \bar b & \bar a
  \end{pmatrix}
\end{equation}
where $a$ and $b$ are complex numbers satisfying $|a|^2 - |b|^2 = 1$.
Similarly $\SU(2)$ is the group of matrices of the form
\begin{equation}
  \label{eq:su2}
  \begin{pmatrix}
    a & b \\ - \bar b & \bar a
  \end{pmatrix}
\end{equation}
where $a$ and $b$ are complex numbers satisfying $|a|^2 + |b|^2 = 1$.
The groups $\SU(1,1)$ and $\SU(2)$ admit bi-invariant metrics
induced from ad-invariant inner products in their respective Lie
algebras:
\begin{equation}
  \label{eq:metric}
  \left<X,Y\right> = \pm \half \Tr XY~,
\end{equation}
where the positive sign is for $X,Y \in \fsu(1,1)$ and the negative
sign for $X,Y \in \fsu(2)$.

Consider the $\U(1)$ subgroup of $\SU(1,1) \times \SU(2)$ consisting
of diagonal matrices
\begin{equation}
  \label{eq:circle}
  \begin{pmatrix}
    e^{i\varphi} &  & & \\
    & e^{-i\varphi} & & \\
    & & e^{i\varphi} &  \\
    & & & e^{-i\varphi} \\
  \end{pmatrix}\qquad\text{where $\varphi \in \RR/2\pi\ZZ$.}
\end{equation}
This subgroup defines a null circle in $\SU(1,1) \times \SU(2)$
relative to the bi-invariant metric.  Indeed, the tangent vector to
the circle at the above element is
\begin{equation*}
  \begin{pmatrix}
    i e^{i\varphi} & & & \\
    & -i e^{-i\varphi} & & \\
    & & i e^{i\varphi} & \\
    & & & -i e^{-i\varphi} \\
  \end{pmatrix}~,
\end{equation*}
whose norm vanishes relative to the inner product defined by
\eqref{eq:metric}.

Let $\sX_i$, $i=0,1,\dots,5$ be the following pseudo-orthonormal basis
for $\fsu(1,1) \oplus \fsu(2)$:
\begin{gather*}
  \sX_0 = i \sigma_3 \oplus 0 \qquad \sX_1 = \sigma_1 \oplus 0 \qquad
  \sX_2 = \sigma_2 \oplus 0\\
  \sX_3 = 0 \oplus i\sigma_1\qquad \sX_4 = 0 \oplus i\sigma_2\qquad \sX_5
  = 0 \oplus i\sigma_3~,
\end{gather*}
where the $\sigma_i$ are the standard (hermitian) Pauli matrices.  The
nonzero Lie brackets are given by
\begin{equation*}
  \begin{aligned}[m]
    [\sX_0,\sX_1] &= -2 \sX_2 \\
    [\sX_0,\sX_2] &= 2 \sX_1 \\
    [\sX_1,\sX_2] &= 2 \sX_0
  \end{aligned}\qquad
  \begin{aligned}[m]
    [\sX_5,\sX_3] &= - 2 \sX_4 \\
    [\sX_5,\sX_4] &= 2 \sX_3 \\
    [\sX_3,\sX_4] &= - 2 \sX_5
  \end{aligned}~,
\end{equation*}
and the inner product is
\begin{equation*}
  \left<\sX_i,\sX_j\right> = \eta_{ij}~,
\end{equation*}
where $\eta = \text{diag}(-1,1,\dots,1)$.

The Lie algebra of the circle subgroup is generated by $\sU = \sX_0 +
\sX_5$.  We see from the above inner product that $\sU$ is indeed null.
Let $\sV = \sX_0 - \sX_5$ be the complementary null generator.  In order
to define the contraction, we let $\Omega > 0$ and introduce new
generators by
\begin{equation*}
  \sP_i = \Omega \sX_i \qquad \sJ = \half \sU  \qquad \sK = \Omega^2
  \sV~,
\end{equation*}
for $i=1,2,3,4$.  The contracted Lie algebra $\fn$ is defined as the
limit $\Omega \to 0$ of the brackets of the new generators:
\begin{equation*}
  \begin{aligned}[m]
    [\sJ,\sP_1] &= - \sP_2\\
    [\sJ,\sP_2] &=  \sP_1
  \end{aligned}\qquad
  \begin{aligned}[m]
    [\sJ,\sP_3] &= - \sP_4\\
    [\sJ,\sP_4] &=  \sP_3
  \end{aligned}\qquad
  \begin{aligned}[m]
    [\sP_1,\sP_2] &= \sK\\
    [\sP_3,\sP_4] &= \sK
  \end{aligned}
\end{equation*}
with $\sK$ central.  The resulting Lie algebra is solvable: indeed its
second derived ideal is central.  It is essentially a Heisenberg Lie
algebra with central element $\sK$, together with an outer
automorphism $\sJ$.  It admits an ad-invariant inner product inherited
from the one in $\fsu(1,1) \oplus \fsu(2)$.  In fact, the inner
product on the generators $\{\sP_i,\sJ,\sK\}$ is given by taking the
limit $\Omega\to 0$ of $\Omega^{-2}\left<-,-\right>$, where
$\left<-,-\right>$ is the inner product in \eqref{eq:metric}.  The
nonzero inner products are
\begin{equation*}
  \left<\sP_i,\sP_j\right> = \delta_{ij} \qquad
  \left<\sJ,\sK\right> = -1~,
\end{equation*}
where $i,j=1,2,3,4$.

We can view this Lie algebra $\fn$ from another point of view, which
is more useful in determining its conjugacy classes.  We can
understand it is a central extension of a subalgebra of the
four-dimensional euclidean algebra.  Indeed, the generators $\sP_i$
are translations in $\RR^4$ and $\sJ$ is a combined rotation in the
$(12)$ and $(34)$ planes.  Together they span a subalgebra of the
euclidean algebra $\fso(4) \ltimes \RR^4$.  Introducing the complex
structure
\begin{equation*}
  J = 
  \begin{pmatrix}
    0 & -1 & 0 & \phantom{-}0\\
    1 & \phantom{-}0 & 0 & \phantom{-}0\\
    0 & \phantom{-}0 & 0 & -1\\
    0 & \phantom{-}0 & 1 & \phantom{-}0
  \end{pmatrix}~,
\end{equation*}
we can rewrite the Lie algebra as
\begin{equation*}
  [\sJ,\sP_i] = \sum_j J_{ij} \sP_j \qquad [\sP_i,\sP_j] = - J_{ij}
  \sK~.
\end{equation*}
In the language of \cite{MedinaRevoy} (see also \cite{FSsug}) this
exhibits the Lie algebra as the \emph{double extension} of the abelian
Lie algebra $\RR^4$ by $\RR$, which explains why this is a solvable
Lie group admitting a bi-invariant inner product.

\subsection{A six-dimensional Nappi--Witten group}

The interpretation of the Lie algebra $\fn$ as a central extension of
a subalgebra of the euclidean algebra makes it very easy to write down
the corresponding Lie group $\eN$.  It will prove convenient to think
of $\RR^4$ with the complex structure $J$ as $\CC^2$.  Indeed, on the
complex linear combinations $\sP_1 + i \sP_2$ and $\sP_3 + i \sP_4$,
the generator $\sJ$ acts by multiplication by $i$.  Let us define
$R(\theta) = \exp(\theta\sJ)$ and $T(u) = \exp(u \cdot \sP)$, where
for $u\in\CC^2$, $u \cdot \sP = \re \bar u^t \sP$, where $\sP = (\sP_1
+ i \sP_2, \sP_3 + i \sP_4)$.  We then have the following group
multiplications
\begin{equation}
  \label{eq:esg}
  \begin{split}
    R(\theta) R(\theta') &= R(\theta + \theta')\\
    R(\theta) T(u) &= T(e^{-i\theta}u) R(\theta)~.
  \end{split}
\end{equation}
The central extension now makes the translation algebra
noncommutative.  Indeed, using the Baker--Campbell--Hausdorff formula
it is easy to find
\begin{equation}
  \label{eq:nct}
  T(u) T(u') = T(u+u') Z(\half \omega(u,u'))~,
\end{equation}
where $Z(t) = \exp(t \sK)$ and where $\omega$ is the symplectic
structure defined by
\begin{equation*}
  \omega(u,u') = \im ( \bar u^t u' )~.
\end{equation*}
Therefore the group elements of (the universal covering group of)
$\eN$ are parametrised by $(u,\theta,t) \in \CC^2 \times \RR \times
\RR$ and the corresponding group element $g(u,\theta,t)$ is given by
\begin{equation*}
  g(u,\theta,t) = T(u) R(\theta) Z(t)~.
\end{equation*}
It is easy to work out the group multiplication law from equations
\eqref{eq:esg} and \eqref{eq:nct}, and one finds:
\begin{equation}
  \label{eq:mult}
  g(u,\theta,t) g(u',\theta',t') = g\left(u + e^{-i\theta} u', \theta +
    \theta', t+ t' + \half \omega(u,e^{-i\theta}u')\right)~.
\end{equation}
It follows that the identity is $g(0,0,0)$ and that the inverse of
$g(u,\theta,t)$ is given by
\begin{equation}
  \label{eq:inverse}
  g(u,\theta,t)^{-1} = g(-e^{i\theta} u, -\theta, -t)~.
\end{equation}

Finally let us work out the bi-invariant metric on the group in this
(global) coordinate system.  The metric at $(u,\theta,t)$ is given by
\begin{equation}
  \label{eq:groupmetric}
  ds^2 = \left< g^{-1}dg, g^{-1}dg\right>~,
\end{equation}
where $g = g(u,\theta,t)$.  A simple calculation shows that
\begin{equation}
  \label{eq:MCL}
  g^{-1}dg(u,\theta,t) = \left(e^{i\theta} du \right) \cdot \sP +
  d\theta \sJ + \left( dt - \half \omega(u,du) \right) \sK~.
\end{equation}
Explicitly, if $u=(u_1 + i u_2, u_3 + i u_4)$, one has
\begin{multline*}
  \left(e^{i\theta} du \right) \cdot \sP = ( du_1 \cos\theta -
  du_2\sin\theta) \sP_1 + (du_2\cos\theta + du_1\sin\theta) \sP_2\\
  + ( du_3 \cos\theta - du_4\sin\theta) \sP_3 + (du_4\cos\theta +
  du_3\sin\theta) \sP_4~.
\end{multline*}
Inserting this into \eqref{eq:groupmetric}, we obtain the following
expression for the bi-invariant metric on the group:
\begin{equation*}
  ds^2 = |du|^2 + \omega(u,du) d\theta - 2 d\theta dt~.
\end{equation*}
Changing coordinates to $x = e^{i\theta/2} u$, $x^+ = -2 t$, $x^-
= \half \theta$, the metric becomes
\begin{equation}
  \label{eq:NW6metric}
  ds^2 = |dx|^2 - |x|^2  (dx^-)^2 + 2 dx^+dx^-~,
\end{equation}
which we recognise as a Cahen--Wallach metric, corresponding to a
(conformally flat) indecomposable lorentzian symmetric space
\cite{CahenWallach}.  In fact, it is a six-dimensional analogue of the
Nappi--Witten spacetime in \cite{NW} and hence will be denoted
$\NW_6$ when we want to emphasise its geometry instead of the
Lie group structure.  As a vacuum solution to the minimal chiral
six-dimensional supergravity, it was discovered by Meessen
\cite{Meessen}.

The fact that a symmetric space is isometric to a Lie group with a
bi-invariant metric is of course not unusual.  After all, every group
$G$ admitting a bi-invariant metric is itself a symmetric space:
simply consider $(G\times G)/G$, where we quotient by the diagonal
subgroup.  However for the Cahen--Wallach symmetric spaces of the type
considered here, the usual description is $G/K$ where $G$ is a
solvable Lie group and $K$ an abelian subgroup.  The question arises
as to whether every Cahen--Wallach space is isometric to some Lie
group with a bi-invariant metric.  Before turning our attention to the
description of symmetric branes for this geometry, let us take a
moment to answer this question, since the answer might be of
independent interest.

\subsection{Symmetric plane waves and lorentzian Lie groups}

It follows from the structure theorem of \cite{MedinaRevoy} and the
refinement in \cite{FSsug,FSalgebra} that indecomposable lorentzian
Lie groups are either simple (e.g., $\SU(1,1)$) or solvable.  In order
to obtain a symmetric plane wave metric we need only concentrate on
the solvable case.  The only solvable Lie groups admitting a
bi-invariant metric are those whose Lie algebras are obtained from the
one-dimensional Lie algebra by iterating two constructions: double
extension and orthogonal direct sum.  Furthermore if the metric is
lorentzian then there is at most one double extension.  Let us focus
on indecomposable Lie algebras admitting a lorentzian invariant
metric.  In this case there is precisely one double extension and in
fact it is not hard to show that these Lie algebras are given by the
double extension of an abelian Lie algebra $\EE^n$ (with the standard
euclidean inner product $\left<-,-\right>$) by the one-dimensional Lie
algebra $\RR$.  The action of $\RR$ on $\EE^n$ is generated by a
skew-symmetric endomorphism $J: \EE^n \to \EE^n$.  If
$\bx,\by\in\EE^n$ and $e_-$ generates $\RR$ and $e_+$ generates the
dual algebra $\RR^*$, then the Lie brackets are given by
\begin{equation*}
  [e_-,\bx] = J(\bx) \qquad\text{and}\qquad [\bx,\by] =
  \omega(\bx,\by) e_+~,
\end{equation*}
where $\omega(\bx,\by) = \left<J(\bx),\by\right>$.  This is a solvable
Lie algebra admitting an invariant scalar product extending
$\left<-,-\right>$ on $\EE^n$ by declaring $\left<e_-,e_+\right>=1$.
The algebra will be indecomposable if $J$ is non-degenerate, which
requires $n$ to be even.  Let $G_\omega$ denote the 1-connected
solvable Lie group with this Lie algebra.  The invariant scalar
product on the Lie algebra induces a bi-invariant metric on
$G_\omega$, which relative to coordinates similar to the ones used
above, takes the Cahen--Wallach form
\begin{equation*}
  ds^2 = 2 dx^+ dx^- - \left<J\bx,J\bx\right> (dx^-)^2 +
  \left<d\bx,d\bx\right>~.
\end{equation*}
Recall that a general Cahen--Wallach metric depends on a symmetric
matrix $A$ in the form
\begin{equation*}
  ds^2 = 2 dx^+ dx^- + A(\bx,\bx) (dx^-)^2 + \left<d\bx,d\bx\right>~.
\end{equation*}
Comparing the two metrics we see that a Cahen--Wallach metric is the
bi-invariant metric on a solvable Lie group if and only if the matrix
$A = J^2$ is negative-definite and every eigenvalue has even
multiplicity.  In particular this means that the metric of the IIB
maximally supersymmetric wave \cite{NewIIB} is in fact a bi-invariant
metric on a solvable Lie group, whereas the metric of the maximally
supersymmetric M-wave \cite{KG} is not.  Of course, strings
propagating in the IIB wave do not really see the Lie group structure
\footnote{although it might see a $4$-Lie group structure
  \cite{FOPPluecker}, if such a thing exists.}  since this background
has no $B$-field, but rather the RR self-dual five-form.

\section{Lie branes in $\SU(1,1) \times \SU(2)$}
\label{sec:LBAdSxS}

The Lie branes in $\SU(1,1) \times \SU(2)$ wrap submanifolds which are
given by (twisted, shifted) conjugacy classes of the group.  These
were analysed originally in \cite{Sads3} and \cite{BPAdS2}.  We
briefly review these results to ease the comparison.

The conjugacy classes of $\SU(2)$ are parametrised by $T/\ZZ_2$, where
$T$ is a maximal torus and $\ZZ_2$ is the Weyl group which acts with
two fixed points.  The quotient is therefore an interval, which we can
take to be $[0,\pi]$.  The conjugacy class corresponding to
$\theta\in[0,\pi]$ is a round 2-sphere with radius $\sin^2\theta$,
hence it degenerates to a point at the endpoints of the interval.  The
induced metric on the two-dimensional spheres is euclidean.  The group
$\SU(2)$ does not have outer automorphisms, hence no twisted
D-branes.

If we parametrise $\SU(1,1)$ as follows
\begin{equation*}
  \SU(1,1) = \left\{ 
    \begin{pmatrix}
      x + i y & u + i v\\ u - i v & x - i y
    \end{pmatrix}
    \biggl|\, x^2 + y^2 = 1 + u^2 + v^2 \right\}~,
\end{equation*}
then conjugacy classes are essentially the intersection of the
hyperboloid $x^2 + y^2 = 1 + u^2 + v^2$ in $\EE^{2,2}$ with the affine
hyperplanes $x=\text{constant}$.  More precisely, when $|x|=1$ each of
the resulting intersections breaks up into three conjugacy classes
corresponding to the apex of a cone and the upper and lower deleted
cones.  Similarly when $|x|<1$ the intersection is a two-sheeted
hyperboloid and each sheet is a conjugacy class.  Keeping only those
conjugacy classes on which the metric is nondegenerate, we have (see
\cite{Sads3} for more details):
\begin{enumerate}
\item 2 pointwise classes corresponding to $\pm \1$;
\item a family of two-dimensional lorentzian submanifolds isometric to
  $\dS_2$ with (squared) radius of curvature proportional to $x^2 -
  1$; and
\item a family of two-dimensional non-flat riemannian submanifolds
  isometric to hyperbolic spaces with (squared) radius of curvature
  proportional to $1-x^2$.
\end{enumerate}
The group $\SU(1,1)$ does have outer automorphisms, giving rise to
twisted conjugacy classes.  Up to inner automorphisms there is a
unique nontrivial outer automorphism which is realised by complex
conjugation in the fundamental representation.  The twisted conjugacy
classes are one-sheeted hyperboloids which can be understood as the
intersection of the hyperboloid $x^2 + y^2 = 1 + u^2 + v^2$ with the
affine hyperplanes $u=\text{constant}$.  The corresponding metric is
lorentzian and has negative constant sectional curvature, whence it is
isometric to an embedded $\AdS_2$ in $\AdS_3$.  These twisted D-branes
were discovered by Bachas and Petropoulos \cite{BPAdS2}.

We can now enumerate the possible Lie branes in $\AdS_3 \times S^3$:
\begin{enumerate}
\item pointlike D-instantons;
\item D-strings sitting at a point in $S^3$, with worldvolume
  geometries $\AdS_2 \subset \AdS_3$ or $\dS_2 \subset \AdS_3$;
\item hyperbolic euclidean D-strings sitting at a point in $S^3$;
\item spherical euclidean D-strings sitting at a point in $\AdS_3$;
\item D3-branes isometric to $\AdS_2 \times S^2$ or $\dS_2 \times
  S^2$; and
\item euclidean D3-branes isometric to $H_2 \times S^2$.
\end{enumerate}

This is now to be compared with the Lie branes in the contracted
group, to which we now turn.

\section{Lie branes in the six-dimensional Nappi--Witten group}
\label{sec:LBNW6}

The problem of determining the submanifolds wrapped by Lie
branes in the six-dimensional Nappi--Witten group $\eN$ obtained by
contracting $\SU(1,1) \times \SU(2)$ consists in determining its
(twisted) conjugacy classes.  As we will see, this problem is very
similar to that for the four-dimensional Nappi--Witten group
\cite{NW}, which was solved in \cite{FSNW}, albeit computationally
somewhat more involved.

\subsection{Orthogonal automorphisms}

The first step is to determine the group of orthogonal automorphisms
of the Nappi--Witten group $\eN$.  Since the exponential map is a
diffeomorphism in this case, we need only determine the orthogonal
automorphisms of its Lie algebra $\fn$.  These are Lie algebra
automorphisms which preserve the metric.  It is straightforward to
show that the most general orthogonal automorphism $\tau:\fn \to \fn$
is given by
\begin{equation}
  \label{eq:autos}
  \begin{split}
    \tau(\sP_i) &= \sum_j M_{ji} \sP_j + \varepsilon \sum_j M_{ji} L_j
    \sK\\
    \tau(\sJ) &= \sum_i L_i \sP_i + \varepsilon \sJ + \half
    \varepsilon \sum_i L_i L_i \sK\\
    \tau(\sK) &= \varepsilon \sK~,
  \end{split}
\end{equation}
where $L_i$ are arbitrary, $\varepsilon = \pm 1$ and $M_{ij}$ satisfy
\begin{equation*}
  \sum_k M_{ki} M_{kj} = \delta_{ij} \qquad \text{and} \qquad
  \sum_{k,\ell} M_{ki} J_{k\ell} M_{\ell j} = \varepsilon J_{ij}~.
\end{equation*}
In other words, $M_{ij}$ are the entries of an orthogonal matrix $M
\in \O(4)$ which preserves the complex structure $J$ up to a sign.  If
$\varepsilon = 1$, then $M$ preserves the complex structure and hence
the induced orientation in $\RR^4$.  This means that $M$ belongs to
the intersection of $\SO(4)$ and $\GL(2,\CC)$ in $\GL(4,\RR)$, which
is isomorphic to $\U(2)$.  If $\varepsilon = -1$, then $M=UX$ where
$U$ belongs to $\U(2)$ and $X$ is any matrix in $\SO(4)$ which obeys
$XJ =-JX$.  One such possibility is $X = \text{diag}(1,-1,1,-1)$,
which belongs to the normaliser of $\U(2)$ in $\SO(4)$. In other
words, conjugation by $X$ defines an outer automorphism $\chi$ of
$\U(2) \subset \SO(4)$ which in fact is none other than complex
conjugation.  We therefore have two types of elements $(L,U,1)$ and
$(L,UX,-1)$, where $U \in \U(2) \subset \SO(4)$ and $L\in\CC^2\cong
\RR^4$.  It is easy to work out the multiplication law for this group:
\begin{equation*}
  (L,M,\varepsilon) \cdot (L',M',\varepsilon') = (ML' + \varepsilon'
  L, MM', \varepsilon\varepsilon')~.
\end{equation*}
This shows that the group $\Aut_o(\fn)$ of orthogonal automorphisms of
$\fn$ (and hence of $\eN$) is isomorphic to
\begin{equation*}
  \Aut_o(\fn) \cong \CC^2 \rtimes \left( \U(2) \rtimes \ZZ_2 \right)~,
\end{equation*}
where $\U(2) \rtimes \ZZ_2$, the principal extension of $\U(2)$ by the
outer automorphism $\chi$, acts on $\CC^2$ as follows.  We embed
$\CC^2 \subset \Aut_o(\fn)$ as $L \mapsto (L,\1,1)$ and the action of
$\U(2) \rtimes \ZZ_2$ is induced by conjugation in $\Aut_o(\fn)$,
whence
\begin{equation*}
 U \cdot L = UL \qquad\text{and}\qquad UX \cdot L = - U \bar L~.
\end{equation*}
Notice that the invariant subgroup consisting of inner automorphisms
is $\CC^2 \rtimes \U(1)$, where $\U(1) \subset \SO(4)$ is the circle
subgroup generated by $J$.  Let us define the group of outer
(orthogonal) automorphisms:
\begin{equation*}
  \Out_o(\fn) = \frac{\Aut_o(\fn)}{\Inn_o(\fn)} \cong \frac{\CC^2
  \rtimes (\U(2) \rtimes \ZZ_2)}{\CC^2 \rtimes \U(1)}~.
\end{equation*}
Notice that every element $(L,M,\varepsilon) \in \Aut_o(\fn)$ can be
decomposed uniquely as
\begin{equation*}
  (L,M,\varepsilon) = (0,M,\varepsilon) (M^{-1}L,\1,1)~,
\end{equation*}
whence modulo the inner automorphisms we can put $L=0$.  In other
words, the group of outer automorphisms is isomorphic to
\begin{equation*}
  \Out_o(\fn) \cong \frac{\U(2) \rtimes \ZZ_2}{\U(1)}~.
\end{equation*}
Further, observe that a given element $(M,\varepsilon) \in \U(2)
\rtimes \ZZ_2$, is such that $M=U$ if $\varepsilon = 1$ or $M=UX$ if
$\varepsilon = -1$, where $U \in \U(2)$.  Given $U\in\U(2)$, we can
write it as $U=S \delta^{1/2}$, where $S\in \SU(2)$ and $\delta = \det
\U$.  This implies that we can write
\begin{align*}
  (U,1) &= (S,1) (\delta^{1/2},1)\\
  (UX,-1) &= (SX,-1) (\bar\delta^{1/2}, 1)~,
\end{align*}
whence modulo inner automorphisms we can always choose $U$ to lie in
$\SU(2)$.  Moreover $U$ and $-U$ are equivalent modulo inner
automorphisms, whence we have that
\begin{equation*}
  \Out_o(\fn) \cong \frac{\SU(2) \rtimes \ZZ_2}{\ZZ_2}~.
\end{equation*}
In other words, $\Out_o(\fn)$ consists of elements of the form $(S,1)$
and $(SX,-1)$, but where $S$ and $-S$ give the same outer
automorphism.  This result is to be contrasted with the
four-dimensional Nappi--Witten group, whose group of outer
(orthogonal) automorphisms is finite of order $2$.

\subsection{Untwisted Lie branes}

The untwisted Lie branes wrap conjugacy classes of the group $\eN$.
To determine these classes we first compute the adjoint action of
$\eN$ on $\eN$.  To determine their geometry we use the fact that at a
point $g_0 \in \eN$, the tangent space $T_{g_0}\eN$ has two natural
subspaces: $T_{g_0}\eC$, which is the tangent space to the conjugacy
class $\eC$ and $T_{g_0}\eZ$, the tangent space to the centraliser
subgroup $\eZ$ of $g_0$: notice that trivially $g_0 \in \eZ$.  It can
be shown that $T_{g_0} \eZ = (T_{g_0}\eC)^\perp$, but it may happen
that $T_{g_0} \eZ \cap T_{g_0}\eC \neq \{0\}$.  This happens if and
only if the restriction of the metric to the conjugacy class $\eC$ (or
equivalently to the centraliser $\eZ$) is degenerate.  In this case,
we cannot interpret the conjugacy class as a D-submanifold, at least
in a straightforward manner, and we will ignore these cases
presently.  If the metric restricts to $\eC$ non-degenerately, then
$T_{g_0} \eN = T_{g_0} \eC \oplus T_{g_0} \eZ$ and the direct sum is
orthogonal.  We can then easily determine the signature of the metric
on the conjugacy class by determining that of the metric on the
centraliser subgroup, which is often easier to determine, since
bi-invariance allows us to work at the identity and hence with the Lie
algebra.

From equations \eqref{eq:mult} and \eqref{eq:inverse} we see that
\begin{multline}
  \label{eq:adjoint}
  g(u,\theta,t) g(u_0,\theta_0,t_0) g(u,\theta,t)^{-1} \\
  = g\left(\left( 1 - e^{-i\theta_0}\right) u + e^{-i\theta} u_0,
  \theta_0, t_0 - \half\im\left( \bar u_0^t e^{i\theta}
  \left(1+e^{-i\theta_0} \right) \right) \right)~,
\end{multline}
from where we deduce that $\theta_0$ is an invariant of the conjugacy
class; whence the induced metric is euclidean, although possibly
degenerate.  Hence we fix a value of $\theta_0$ and investigate the
action on the remaining coordinates $(u_0,t_0)$.  Furthermore,
periodicity allows us to restrict our attention to $\theta_0 \in
\RR/2\pi\ZZ$.

\subsubsection{$\theta_0 = 0 \pmod{2\pi}$}  Conjugation maps
\begin{equation*}
  (u_0,t_0) \mapsto \left(e^{-i\theta} u_0, t_0 - \im\left(\bar u_0^t
  e^{i\theta} u\right)\right)~.
\end{equation*}
If $u_0 = 0$, then $g(0,0,t_0)$ belongs to the centre and hence the
conjugacy class is a point.  If $u_0 \neq 0$ we obtain the cylindrical
conjugacy class consisting of elements
\begin{equation*}
  g(e^{-i\theta} u_0, 0, t)\qquad \text{for all $t\in\RR$ and
  $\theta\in\RR/2\pi\ZZ$,}
\end{equation*}
which is diffeomorphic to $\RR \times S^1$.  A routine calculation
shows that the metric restricts degenerately, whence these cylindrical
conjugacy classes cannot be straightforwardly interpreted as
D-branes.

\subsubsection{$\theta_0 = \pi \pmod{2\pi}$}  Conjugation maps
\begin{equation*}
  (u_0,t_0) \mapsto (2u + e^{-i\theta} u_0, t_0)~,
\end{equation*}
which is the $4$-plane labelled by $t_0$.  It is easy to see that the
induced metric has euclidean signature.  Thus these conjugacy classes
are wrapped by euclidean D3-branes.

\subsubsection{$\theta_0 \neq 0,\pi \pmod{2\pi}$}  We first determine
the centraliser of an element $g(u_0,\theta_0,t_0)$.  We find that it
consists of the elements
\begin{equation*}
  g(u(\theta), \theta, t) \quad\text{where}\quad u(\theta) =
  \frac{1-e^{-i\theta}}{1-e^{-i\theta_0}} u_0~.
\end{equation*}
Since the centraliser is two-dimensional (parametrised by $\theta$ and
$t$), the conjugacy classes are four-dimensional.  The metric on the
centraliser is lorentzian, whence the metric on the conjugacy class at
the chosen element (and by homogeneity everywhere) is euclidean,
whence these conjugacy classes are wrapped by euclidean D3-branes.

In summary, the conjugacy classes can be wrapped by D-instantons and
euclidean D3-branes.

\subsection{Twisted Lie branes}

We now turn our attention to the twisted conjugacy classes
\begin{equation*}
  C^r(g_0) = \left\{ r(g) g_0 g^{-1} | g \in \eN \right \}
\end{equation*}
where $r$ is an orthogonal automorphism.  As discussed above,
the orthogonal outer automorphism group consists of elements
of the form $(M,\varepsilon)$, where $M= S$ if $\varepsilon=1$ and
$M=SX$ if $\varepsilon = -1$, where $S\in\SU(2)$.  We repeat that $S$
and $-S$ give rise to the same outer automorphism.  If
$g=g(z,\theta,t)$ is a group element and $r$ is the automorphism
corresponding to $(M,\varepsilon)$, then
\begin{equation*}
  r(g) = g(Mz, \varepsilon \theta, \varepsilon t) = 
  \begin{cases}
    g(Sz, \theta, t) & \text{if $\varepsilon=1$}\\
    g(S\bar z, -\theta, -t) & \text{if $\varepsilon=-1$.}
  \end{cases}
\end{equation*}
Under twisted conjugation by $g$, the element
$g_0=g(z_0,\theta_0,t_0)$ is mapped to
\begin{equation*}
  r(g) g_0 g^{-1} = g(z',\theta',t')~,
\end{equation*}
where, if $\varepsilon=1$,
\begin{equation}
  \label{eq:orbit+}
  \begin{aligned}
    z' &= Sz + e^{-i\theta} z_0 - e^{-i\theta_0} z\\
    \theta' &= \theta_0\\
    t' &= t_0 + \half \omega(Sz, e^{-i\theta}z_0) - \half \omega(Sz +
    e^{-i\theta}z_0, e^{-i\theta_0} z)~;
  \end{aligned}
\end{equation}
whereas, if $\varepsilon = -1$,
\begin{equation}
  \label{eq:orbit-}
  \begin{aligned}
    z' &= S\bar z + e^{i\theta} z_0 - e^{-i(\theta_0-2\theta)} z\\
    \theta' &= \theta_0 - 2 \theta\\
    t' &= t_0 - 2 t + \half \omega(S\bar z, e^{i\theta}z_0) - \half
    \omega(S\bar z + e^{i\theta}z_0, e^{-i(\theta_0-2\theta)} z)~.
  \end{aligned}
\end{equation}

\subsubsection{The case $\varepsilon=1$}

Let us first consider the case of $\varepsilon=1$.  We notice that
$\theta_0$ is an orbit invariant, whence the metric on the orbit will
be euclidean, but perhaps degenerate.  The stabiliser of $g_0$ is the
subgroup consisting of elements of the form $g(z,\theta,t)$ where $t$
is unconstrained, and where $z,\theta$ satisfy
\begin{equation}
  \label{eq:stab}
  \begin{aligned}
    (S-e^{-i\theta_0})z &= (1 - e^{-i\theta})z_0\\
    \omega(Sz, e^{-i\theta} z_0) &= \omega(Sz + e^{-i\theta}z_0,
    e^{-i\theta_0} z)~.
  \end{aligned}
\end{equation}
Inserting the first equation into the second, allows us to rewrite it
as
\begin{equation}
  \label{eq:stab2}
  \omega(e^{-i\theta_0}z + z_0, (1 + e^{-i\theta})z_0) = 0~.
\end{equation}
Two cases present themselves according to whether $z_0$ vanishes or
not.

If $z_0=0$ then the second equation is automatically satisfied,
$\theta$ is unconstrained and $z$ satisfies
\begin{equation*}
  (S - e^{-i\theta_0})z=0~.
\end{equation*}
Therefore the resulting stabiliser consists of elements
$g(z,\theta,t)$ where $z$ obeys the above equation.  We can break this
up further into three cases depending on the rank of the complex
linear map $\varphi := S - e^{-i\theta_0}$.  In the generic case,
$\varphi$ will have (complex) rank $2$ and hence $z=0$.  The resulting
stabiliser consists of $g(0,\theta,t)$ which is two-dimensional,
whence the orbit is four-dimensional: a euclidean D$3$-brane.  If
$\varphi$ has rank $1$, so that $e^{i2\theta_0} \neq 1$, then the
stabiliser is four-dimensional, and hence the orbit is a euclidean
D-string.  Finally the case where $\varphi$ has zero rank, which can
only happen when $e^{i2\theta_0} = 1$ and $S = \pm \1$, reduces to an
inner automorphism. This has been considered in the previous section
and one obtains a D-instanton.

Now suppose that $z_0\neq 0$.  We again distinguish several cases
depending on the rank of $\varphi$, although now $z$ is not
necessarily in its kernel.  In the generic case of rank $2$, then
$\varphi$ is invertible and we can solve for $z$:
\begin{equation*}
  z = \varphi^{-1} (1-e^{-i\theta}) z_0~.
\end{equation*}
Inserting this into the second equation one finds that (remarkably?)
the equation is automatically satisfied for any value of $\theta$.
Therefore the stabiliser, which consists of elements
$g(z(\theta),\theta,t)$, is clearly two-dimensional and hence, for
generic $\theta_0$, the twisted conjugacy classes are wrapped by
euclidean D3-branes.

There are two types of non-generic $\theta_0$: depending on whether
the (complex) rank of $\varphi$ is $1$ or $0$.  This latter
case can only happen when $S=\pm\1$ and $\theta_0 =k\pi$ for
$k\in\ZZ$.  This latter case corresponds to twisting by an inner
automorphism, whence it was discussed in the previous section.  We
will therefore focus briefly on the case of $\varphi$ having rank $1$,
and $S\neq \pm \1$.

There are two cases that we must consider, depending on whether or not
$z_0$ belongs to the image of $\varphi$.  If $z_0
\not\in\image\varphi$ then the first equation in
\eqref{eq:stab} can only be satisfied if $e^{i\theta} =1$ and
$z\in\ker\varphi$.  The second equation $\omega(z,e^{i\theta_0}z_0)=0$
either gives none or one (real) equation on $z$, so that the
stabiliser subgroup is either three-dimensional or two-dimensional
respectively.  The resulting Lie branes are then euclidean D2- or
D3-branes, provided the metric restricts non-degenerately.  We have
not checked, since we do not think they are particularly interesting.

Finally let us consider the case where $z_0 = \varphi(w_0)$ for
some $w_0$.  In this case $z = (1-e^{-i\theta})w_0 + w$, where $w \in
\ker \varphi$.  Since $z_0\neq 0$, $\varphi$ has rank $1$ or $2$.
The case of full rank has $z=z(\theta)$ and then the second equation
in \eqref{eq:stab} fixes $\theta$.  As a result the stabiliser is
one-dimensional, whence the orbit if five-dimensional but inherits a
degenerate metric.  If the rank of $\varphi$ is not maximal, then the
stabiliser has dimension three and, provided the metric is not
degenerate (which we have not checked) the orbit can be wrapped by
euclidean D2-branes.

In summary, all the Lie branes wrapping twisted conjugacy classes
associated to outer automorphisms with $\varepsilon = 1$ are
euclidean: D-instantons, D-strings, D2-branes (possibly) and
D3-branes.

\subsubsection{The case $\varepsilon=-1$}
\label{sec:D5}

From the explicit expression \eqref{eq:orbit-} for the twisted
conjugacy of an element $g(z_0,\theta_0,t_0)$, we find that the
stabiliser subgroup consists of elements $g(z,0,t)$ where\footnote{A
  word of caution: we have set $\theta=0$ because we are working in
  the universal cover of the group, so that $\theta_0 \in \RR$. If
  instead we work in a quotient where $\theta_0 \in \RR/2\pi n\ZZ$,
  then the condition is that $\theta \in \pi n \ZZ$, so that
  $e^{i\theta} = (-1)^n$, whence for odd $n$ we would have to modify
  the discussion below.}
\begin{equation*}
  4 t = \omega(S\bar z, z_0) - \omega(S\bar z + z_0, e^{-i\theta_0} z)~,
\end{equation*}
and $z$ satisfies the \emph{real} linear equation
\begin{equation*}
  S \bar z = e^{-i\theta_0} z~.
\end{equation*}
Inserting this equation into the above expression for $t$ we find
\begin{equation*}
  t = \half \omega(e^{-i\theta_0}z, z_0)~.
\end{equation*}
The dimension of the orbit depends the rank of the real linear map
$\varrho: \RR^4 \to \RR^4$ defined by $z \mapsto S \bar z -
e^{-i\theta_0}z$, where $S \in \SU(2)$.  A straightforward calculation
shows that $\varrho$ is invertible \emph{unless} the off-diagonal
entries of $S$ are pure imaginary.  Concretely, if $S =
\begin{pmatrix} a & b \\ -\bar b & \bar a \end{pmatrix}$, then $\det
\varrho = 4 (\im b)^2$.  Moreover if $\im b =0$ then the (real) rank
of $\varrho$ is $2$.  Therefore, for generic $S$, the stabiliser is
trivial and the resulting Lie brane is a space-filling D5-brane.  If
$S$ is such that $\varrho$ has rank $2$ then the resulting Lie brane
is a D3-brane.  We will see below that these D3-branes have
the geometry of a four-dimensional Nappi--Witten spacetime.  We see
thus that the six-dimensional Nappi--Witten group admits many
foliations by four-dimensional Nappi--Witten spacetimes.

To summarise the results of this section, the Lie branes of $\NW_6$
are of the following types: D-instantons, euclidean D-strings,
euclidean D2-branes (possibly), euclidean D3-branes and perhaps the
most interesting ones are lorentzian D3-branes and space-filling
D5-branes.

\subsection{Geometry of lorentzian Lie branes}
\label{sec:D3}

We will focus on the Lie branes with lorentzian signature and
investigate their geometry.  The space-filling D5-branes are
of course isometric to the spacetime itself; hence
we will concentrate on the D3-branes.  The geometry of the D3-branes
can be worked out from the formulae for the metric and the embedding
of the twisted conjugacy classes, once we find a convenient
parametrisation.  We will show that the D-submanifolds wrapped by the
D3-branes are isometric to the Nappi--Witten spacetime \cite{NW}.

This is not unexpected, based on the analogous analysis of the Lie
branes of the Nappi--Witten spacetime itself \cite{FSNW}.  Indeed the
twisted conjugacy classes of $\NW_4$, whose metrics are
given in equations (22) and (23) of \cite{FSNW}, are themselves
three-dimensional symmetric plane waves.  Indeed they are isometric to
the unique three-dimensional symmetric plane wave, as can be seen by
changing variables in those equations to bring both metrics to the
form
\begin{equation*}
  ds^2 = dx^2 - x^2 (dx^-)^2 + 2dx^+ dx^-~.
\end{equation*}

We can also argue purely geometrically that D-submanifold isometric to
$\NW_4$ exists in $\NW_6$.  Indeed, as shown in \cite{Limits}, the
Nappi--Witten metric arises as a Penrose limit of $\AdS_2 \times S^2$
(with equal radii of curvature) along any null geodesic having a
nonzero velocity component tangent to the sphere.\footnote{This is not
  the only way this metric arises as a Penrose limit: in \cite{GONW4}
  it arises as the Penrose limit of the near-horizon geometry of an
  NS5-brane.} On the other hand, we know that $\AdS_3 \times S^3$ has
a family of D-branes with the geometry of $\AdS_2 \times S^2$.  We
would like to find a situation in which the Penrose limit will
simultaneously induce the Penrose limits of the brane and of the
ambient spacetime.  In other words, we would like to encounter a
situation where the following diagram commutes
\begin{equation}
  \begin{CD}
    @. \\
    \AdS_3 \times S^3 @>\text{Penrose limit}>> \NW_6  \\
    @AAA          @AAA \\
    \AdS_2 \times S^2 @>\text{Penrose limit}>> \NW_4\\
    @.
  \end{CD}
\end{equation}
where the vertical arrows are isometric embeddings given by the
corresponding twisted conjugacy classes.

To achieve this, it is enough to ensure the existence a null geodesic
in $\AdS_3 \times S^3$ (with nonzero velocity component tangent to the
sphere) which starts from a point in one of these $\AdS_2 \times S^2$
and whose initial velocity is tangent to the $\AdS_2 \times S^2$
\emph{and} which stays on $\AdS_2 \times S^2$ for all time.  Then the
Penrose limit of $\AdS_3 \times S^3$ along this null geodesic, which
gives rise to $\NW_6$, induces simultaneously the Penrose limit of
$\AdS_2 \times S^2$, which is known to give rise to the Nappi--Witten
spacetime $\NW_4$.  The above conditions on the null geodesic will be
met if $\AdS_2 \times S^2 \subset \AdS_3 \times S^3$ is a
\emph{totally geodesic} submanifold; although this may be too strong,
since we only require this property for a particular null geodesic.
Let us recall that a submanifold is totally geodesic if and only if
its second fundamental form vanishes.  In principle one could compute
the second fundamental forms of these (twisted) conjugacy classes and
see that there is a totally geodesic $\AdS_2 \times S^2$ among the
twisted D-branes of $\AdS_3 \times S^3$; but in fact we will arrive at
this result by more geometric means.

To this end we employ the useful fact that a hypersurface which is the
fixed point set of a ``reflection'' is automatically totally geodesic.
Let us elaborate on this.  Suppose that $(M,g)$ is a
(pseudo-)riemannian manifold and that $\tau:M\to M$ is an isometry
having as fixed point set a hypersurface $N \subset M$; that is, a
submanifold of codimension one.  We will assume that $g$ restricts
non-degenerately on $N$, so that for every $p \in N$, $T_pM = T_pN
\oplus T_pN^\perp$, and that the derivative map $\tau_*:T_pM \to T_pM$
is a reflection in $T_pN$.  In other words, we have that
\begin{equation*}
  \tau_*|_{T_pN} = \id \qquad\text{and}\qquad 
  \tau_*|_{T_pN^\perp} = -\id~.
\end{equation*}
(This defines what we meant by ``reflection'' above.)  The second
fundamental form of $N \subset M$ defines for every $p \in N$ a
symmetric bilinear map $\II_p: T_pN \times T_pN \to T_pN^\perp$.  The
action of $\tau_*$ is such that $\II_p$ changes sign, but on the other
hand $\tau$ is an isometry which leaves $N$ pointwise fixed, hence
$\II_p$ is invariant.  The only way these two statements can be
reconciled is if $\II_p$ vanishes identically.

We will now exhibit a totally geodesic $\AdS_2\times S^2$ in $\AdS_3
\times S^3$ by exhibiting the $\AdS_2$ and the $S^2$ as fixed point
sets of a reflection on $\AdS_3$ and $S^3$, respectively.  Let us
first consider the $S^2 \subset S^3$ consisting of an equatorial
sphere; in other words, if $S^3$ is defined as the quadric in $\EE^4$
given by
\begin{equation*}
  x^2 + y^2 + z^2 + w^2 = R^2~,
\end{equation*}
then consider the reflection $x \mapsto -x$ and leaving $y,z,w$
invariant.  This is an isometry of $\EE^4$ which preserves and hence
induces an isometry in $S^3$.  This isometry fixes the intersection of
$S^3$ with the hyperplane $x=0$, which is the equatorial two-sphere
given by $x=0$ and $y^2 + z^2 + w^2 = R^2$, which is then totally
geodesic in $S^3$.

Similarly consider the following $\AdS_2 \subset \AdS_3$.  If we think
of $\AdS_3$ as the quadric
\begin{equation*}
  x^2 + y^2 = R^2 + u^2 + v^2 \qquad\text{in $\EE^{2,2}$,}
\end{equation*}
then the family of twisted conjugacy classes isometric to $\AdS_2$ are
those corresponding to $u=\text{constant}$.  Consider the ``time
reversal'' isometry of $\EE^{2,2}$ given by $u\mapsto -u$ and leaving
$x,y,v$ invariant.  This isometry preserves and hence induces an
isometry of $\AdS_3$ which fixes the intersection of $\AdS_3$ with the
hyperplane $u=0$.  The corresponding $\AdS_2$ is therefore totally
geodesic.  Notice that both the totally geodesic $S^2$ and $\AdS_2$
inherit their radius of curvature from the ambient $S^3$ and $\AdS_3$,
respectively; whence if the radius of curvature of the ambient spaces
are equal, so are the ones of the $\AdS_2$ and $S^2$.  In summary we
have exhibited a totally geodesic $\AdS_2 \times S^2$ with equal radii
of curvature inside $\AdS_3 \times S^3$, and hence by the
commutativity of the above diagram, the Penrose limit will give rise
to a $\NW_4$ submanifold of $\NW_6$.

We can exhibit this submanifold explicitly as a twisted conjugacy
class of the Nappi--Witten group $\eN$, as follows.  Consider the
automorphism $r$ of $\eN$ given by
\begin{equation*}
  g(z,\theta,t) \mapsto g(\bar z, -\theta, -t)~;
\end{equation*}
in other words, this is the element $(X,-1)$ in the outer
automorphism group, corresponding to $S=\1$.  Let $g_0 = g(z_0,
\theta_0, t_0)$ be an arbitrary element and consider its orbit under
the twisted adjoint action $r(g) g_0 g^{-1}$, whose elements are given
by equation \eqref{eq:orbit-} with $S=\1$.  Without loss of generality
we parametrise the twisted conjugacy class as
\begin{equation*}
  g(e^{-i\phi/2} e^{i\theta_0/2} z_0 + \bar z - e^{-i\phi}z, \phi,
  s)~,
\end{equation*}
where $\phi,s\in\RR$.  As $z\in\CC^2$ varies, the points
\begin{equation*}
  e^{-i\phi/2} e^{i\theta_0/2} z_0 + \bar z - e^{-i\phi}z
\end{equation*}
define an affine (real two-dimensional) plane through $e^{-i\phi/2}
e^{i\theta_0/2} z_0$ in $\RR^4$ which, after a moment's thought, can
be seen to be parametrised by
\begin{equation*}
  e^{-i\phi/2} (e^{i\theta_0/2} z_0 - 2i y)\qquad\text{for $y \in \RR^2
  \subset \CC^2$.}
\end{equation*}
In summary, the twisted conjugacy class consists of elements
\begin{equation*}
  g(e^{-i\phi/2} (e^{i\theta_0/2} z_0 - 2i y), \phi, s)
\end{equation*}
where $(y,\phi,t)\in\RR^4$.  Write $e^{i\theta_0/2} z_0 = x_0 + i
y_0$, with $x_0,y_0\in\RR^2$.  Define the following coordinates
\begin{equation*}
  x = y_0 - 2y~,\quad x^- = \half \phi \quad\text{and}\quad x^+ = -
  2 s + \tfrac14 |x_0|^2 \phi~,
\end{equation*}
where here and below $|~|^2 $ indicates the euclidean norm of a
two-vector.  Relative to these coordinates the metric induced from
\eqref{eq:NW6metric} takes the form
\begin{equation}
  \label{eq:NW4metric}
  ds^2 = |dx|^2 - |x|^2 (dx^-)^2 + 2 dx^+ dx^-~,
\end{equation}
which we recognise as the metric of the Nappi--Witten spacetime.  In
summary, we have exhibited a foliation of the Penrose limit of $\AdS_3
\times S^3$ consisting of Nappi--Witten spacetimes.  Moreover (at
least some of) these ``braneworlds'' are themselves the Penrose limits
of $\AdS_2 \times S^2$ braneworlds in $\AdS_3 \times S^3$.

How about the other lorentzian D3-branes?  In fact, it is not hard to
show that they are also isometric to Nappi--Witten spacetimes.  To see
this let $S = \begin{pmatrix} a & b \\ -\bar b & \bar a
\end{pmatrix}$, where $b = i \gamma$ is pure imaginary and $a =
\alpha + i \beta$, where $\alpha^2 + \beta^2 + \gamma^2 = 1$.  We will
assume that $\alpha < 1$, for otherwise $S = \1$ and this was the Lie
brane we just described.  The corresponding twisted conjugacy class
consists of points
\begin{equation*}
  g(e^{-i\phi/2} e^{i\theta_0/2} z_0 +  S\bar z - e^{-i\phi} z, \phi, s)~,
\end{equation*}
with $\phi,s\in\RR$.  It is convenient to define $w = e^{-i\phi/2} z$
and $w_0 = e^{i\theta_0/2} z_0$, so that the conjugacy class now
consists of the points
\begin{equation*}
  g(e^{-i\phi/2} (w_0 +  S\bar w - w), \phi, s)~.
\end{equation*}
Our assumption that $\alpha \neq 1$ implies that we can take
$w\in\RR^2 \subset \CC^2$ and this still parametrises the conjugacy
class.  (For $\alpha = 1$ we would (and did) take $iw \in \RR^2$.)
The induced metric on this submanifold is given by
\begin{equation*}
  ds^2 = |S dw - dw|^2 - |w_0 + Sw - w|^2 d\phi^2 - 2d\phi ds~,
\end{equation*}
which can be rewritten as
\begin{equation*}
  ds^2 = 2 (1-\alpha) |dw|^2 - ( 2 (1-\alpha) |w - c|^2 + \mu) d\phi^2
  - 2d\phi ds~,
\end{equation*}
for some $c\in\RR^2$ and $\mu\in\RR$ whose explicit expressions are of
no relevance.  Now simply define the new coordinates
\begin{equation*}
  x = \sqrt{2(1-\alpha)} (w - c)~, \quad x^- = \half\phi
  \quad\text{and}\quad x^+ = -2s - \mu \phi~,
\end{equation*}
relative to which the metric adopts the standard form
\eqref{eq:NW4metric} for the Nappi--Witten spacetime.

\section{Summary and discussion}

In summary we have determined the Lie branes of the symmetric plane
wave $\NW_6$ which arises as the Penrose limit of $\AdS_3 \times S^3$
along a generic null geodesic.  Among the lorentzian Lie branes we
have found D3-branes which are isometric to Nappi--Witten spacetimes
$\NW_4$.  Indeed, $\NW_6$ can be foliated by $\NW_4$ in a variety of
ways.  We have argued that at least one of these D3-branes arises as
the Penrose limit of a totally geodesic $\AdS_2 \times S^2$ Lie brane
of $\AdS_3 \times S^3$.  We also exhibited the whole $\NW_6$ spacetime
as a space-filling D5-brane, but since there are no space-filling Lie
branes in $\AdS_3 \times S^3$, its origin is more mysterious; although
as shown in \cite{KNSD5}, when warped into a fivebrane solution of
type IIB supergravity it can be understood as the Penrose limit of a
(non Lie) fivebrane solution with worldvolume $\AdS_3 \times S^3$.

One can ask about the fate of the other Lie branes in $\AdS_3 \times
S^3$ under the Penrose limit.  If the null geodesic along which the
limit is taken does not lie along the Lie brane, there is not much
that can be said.  Some of the D-strings in $\AdS_3 \times S^3$, e.g.,
those with $\AdS_2$ geometry, behave nicely under the Penrose limit of
$\AdS_3 \times S^3$ along a null geodesic with zero velocity component
along $S^3$, which yields flat space.  These D-strings become
D-strings with flat worldsheets in Minkowski spacetime.  Notice that
if we think of Minkowski spacetime as an abelian Lie group, then these
Lie branes are precisely the twisted conjugacy classes which, when
written additively, are seen to be precisely the affine planes
consisting of the points
\begin{equation*}
  x_0 + r(x) - x
\end{equation*}
where $r$ is now a Lorentz transformation.  Similarly, it is tempting
to conjecture that lorentzian twisted conjugacy classes of lorentzian
Lie groups which are also symmetric plane waves are themselves
symmetric plane waves.

\section*{Acknowledgments}

The research of SS was funded by PPARC under grant SPG/00613, whereas
that of JMF is supported in part by EPSRC under grant GR/R62694/01.
JMF is a member of EDGE, Research Training Network HPRN-CT-2000-00101,
supported by The European Human Potential Programme.  JMF would like
to thank the Rutgers NHETC for hospitality during the final stages of
the writing of this paper.

\bibliographystyle{utphys}
\bibliography{AdS3,ESYM,Sugra,Geometry,CaliGeo}

\providecommand{\href}[2]{#2}\begingroup\raggedright\begin{thebibliography}{10}

\bibitem{PenrosePlaneWave}
R.~Penrose, ``Any space-time has a plane wave as a limit,'' in {\em
  Differential geometry and relativity}, pp.~271--275.
\newblock Reidel, Dordrecht, 1976.

\bibitem{GuevenPlaneWave}
R.~Güven, ``Plane wave limits and {T}-duality,'' {\em Phys. Lett.} {\bf B482}
  (2000) 255--263. \texttt{arXiv:hep-th/0005061}.

\bibitem{MaldaPL}
D.~Berenstein, J.~Maldacena, and H.~Nastase, ``Strings in flat space and pp
  waves from ${N}{=}4$ {S}uper {Y}ang {M}ills.'' \texttt{arXiv:hep-th/0202021}.

\bibitem{Limits}
M.~Blau, J.~Figueroa-O'Farrill, and G.~Papadopoulos, ``Penrose limits,
  supergravity and brane dynamics,'' {\em Class. Quant. Grav.} {\bf 19} (2002)
  4753--4805. \texttt{arXiv:hep-th/0202111}.

\bibitem{NW}
C.~Nappi and E.~Witten, ``A {WZW} model based on a non-semi-simple group,''
  {\em Phys. Rev. Lett.} {\bf 71} (1993) 3751--3753.
  \texttt{arXiv:hep-th/9310112}.

\bibitem{Meessen}
P.~Meessen, ``A small note on pp-wave vacua in 6 and 5 dimensions.''
  \texttt{arXiv:hep-th/0111031}.

\bibitem{InonuWigner}
E.~In\"on\"u and E.~Wigner, ``On the contraction of groups and their
  representations,'' {\em Proc. Nat. Acad. Sci. USA} {\bf 39} (1956) 510--524.

\bibitem{AS}
A.~Alekseev and V.~Schomerus, ``D-branes in the {WZW} model,'' {\em Phys. Rev.}
  {\bf D60} (1999) 061901. \texttt{arXiv:hep-th/9812193}.

\bibitem{FFFS}
G.~Felder, J.~Fr\"ohlich, J.~Fuchs, and C.~Schweigert, ``The geometry of {WZW}
  branes,'' {\em J. Geom. Phys.} {\bf 34} (2000) 162--190.
  \texttt{arXiv:hep-th/9909030}.

\bibitem{SDnotes}
S.~Stanciu, ``D-branes in group manifolds,'' {\em J. High Energy Phys.} {\bf
  01} (2000) 025. \texttt{arXiv:hep-th/9909163}.

\bibitem{FSNW}
J.~Figueroa-O'Farrill and S.~Stanciu, ``More {D}-branes in the {N}appi-{W}itten
  background,'' {\em J. High Energy Phys.} {\bf 01} (2000) 024.
  \texttt{arXiv:hep-th/9909164}.

\bibitem{Sads3}
S.~Stanciu, ``D-branes in an $\text{AdS}_3$ background,'' {\em J. High Energy
  Phys.} {\bf 09} (1999) 028. \texttt{arXiv:hep-th/9901122}.

\bibitem{BPAdS2}
C.~Bachas and M.~Petropoulos, ``Anti-de {S}itter {D}-branes,'' {\em J. High
  Energy Phys.} {\bf 02} (2001) 025. \texttt{arXiv:hep-th/0012234}.

\bibitem{BKPDpp}
A.~Biswas, A.~Kumar, and K.~Panigrahi, ``$p$-$p'$ branes in pp-wave
  background,'' {\em Phys. Rev.} {\bf D66} (2002) 126002.
  \texttt{arXiv:hep-th/0208042}.

\bibitem{NayakDangles}
R.~Nayak, ``D-branes at angle in pp-wave background.''
  \texttt{arXiv:hep-th/0210230}.

\bibitem{KNSD5}
A.~Kumar, R.~Nayak, and Sanjay, ``D-brane solutions in pp-wave background,''
  {\em Phys. Lett.} {\bf B541} (2002) 183--188. \texttt{arXiv:hep-th/0204025}.

\bibitem{Singh}
H.~Singh, ``M5-branes with $3/8$ supersymmetry in pp-wave background.''
  \texttt{arXiv:hep-th/0205020}.

\bibitem{AKD3}
M.~Alishahiha and A.~Kumar, ``D-brane solutions from new isometries of
  pp-waves,'' {\em Phys. Lett.} {\bf B542} (2002) 130--136.
  \texttt{arXiv:hep-th/0205134}.

\bibitem{GONW4}
J.~Gomis and H.~Ooguri, ``Penrose limit of {$N{=}1$} gauge theories,'' {\em
  Nuc. Phys.} {\bf B635} (2002) 106--126,
  \href{http://www.arXiv.org/abs/\texttt{arXiv:hep-th/0202157}}{{\tt
  \texttt{arXiv:hep-th/0202157}}}.

\bibitem{ShortLimits}
M.~Blau, J.~Figueroa-O'Farrill, C.~Hull, and G.~Papadopoulos, ``Penrose limits
  and maximal supersymmetry,'' {\em Class. Quant. Grav.} {\bf 19} (2002)
  L87--L95. \texttt{arXiv:hep-th/0201081}.

\bibitem{ORS}
D.~Olive, E.~Rabinovici, and A.~Schwimmer, ``A class of string backgrounds as a
  semiclassical limit of {WZW} models,'' {\em Phys. Lett.} {\bf B321} (1994)
  361--364. \texttt{arXiv:hep-th/9311081}.

\bibitem{MedinaRevoy}
A.~Medina and P.~Revoy, ``Algèbres de {L}ie et produit scalaire invariant,''
  {\em Ann. scient. Éc. Norm. Sup.} {\bf 18} (1985) 553.

\bibitem{FSsug}
J.~Figueroa-O'Farrill and S.~Stanciu, ``Nonsemisimple {Sugawara}
  constructions,'' {\em Phys. Lett.} {\bf B327} (1994) 40--46.
  \texttt{arXiv:hep-th/9402035}.

\bibitem{CahenWallach}
M.~Cahen and N.~Wallach, ``Lorentzian symmetric spaces,'' {\em Bull. Am. Math.
  Soc.} {\bf 76} (1970) 585--591.

\bibitem{FSalgebra}
J.~Figueroa-O'Farrill and S.~Stanciu, ``On the structure of symmetric selfdual
  {L}ie algebras,'' {\em J. Math. Phys.} {\bf 37} (1996) 4121--4134.
  \texttt{arXiv:hep-th/9506152}.

\bibitem{NewIIB}
M.~Blau, J.~Figueroa-O'Farrill, C.~Hull, and G.~Papadopoulos, ``A new maximally
  supersymmetric background of type {IIB} superstring theory,'' {\em J. High
  Energy Phys.} {\bf 01} (2002) 047. \texttt{arXiv:hep-th/0110242}.

\bibitem{KG}
J.~Kowalski-Glikman, ``Vacuum states in supersymmetric {K}aluza-{K}lein
  theory,'' {\em Phys. Lett.} {\bf 134B} (1984) 194--196.

\bibitem{FOPPluecker}
J.~Figueroa-O'Farrill and G.~Papadopoulos, ``{P}lücker-type relations for
  orthogonal planes.'' \texttt{arXiv:math.AG/0211170}.

\end{thebibliography}\endgroup

\end{document}